\documentclass[twocolumn]{aastex62}
\usepackage[utf8]{inputenc}
\usepackage{longtable}
\usepackage{subfigure}
\usepackage{natbib}
\usepackage{textcomp}
\usepackage[colorinlistoftodos]{todonotes}
\usepackage{multirow}
\usepackage{enumerate}
\usepackage{enumitem}

\newcommand{\code}[1]{\texttt{#1}}
\newcommand{\galpy}{\code{\texttt{galpy}}}

\graphicspath{{./}{figures/}}

\shorttitle{High- and Low-$\alpha$ Disk Stars Separate Dynamically at all Ages}
\shortauthors{Gandhi et al.}
\begin{document}

\title{High- and Low-$\alpha$ Disk Stars Separate Dynamically at all Ages}

\correspondingauthor{Suroor S Gandhi}
\email{ssg487@nyu.edu}

\author{Suroor S Gandhi}
\affil{Department of Physics, New York University,
726 Broadway,
New York, NY, 10003, USA}

\author{Melissa K Ness}
\affiliation{Center for Computational Astrophysics, Flatiron Institute,
162 5th Ave., New York, NY 10010, USA}
\affiliation{Department of Astronomy, Columbia University, 550 W
120th St, New York, NY 10027, USA}
% \collaboration{(AAS Journals Data Scientists collaboration)}



\begin{abstract}
There is a dichotomy in the Milky Way in the $[\alpha/$Fe]-[Fe/H] plane, in which stars fall into high-$\alpha$, and low-$\alpha$ sequences. The high-$\alpha$ sequence comprises mostly old stars, and the low-$\alpha$ sequence comprises primarily young stars. The origin of this dichotomy  is uncertain. To better understand how the high- and low-$\alpha$-stars are affiliated, we examine if the high- and low-$\alpha$ sequences have distinct orbits at all ages, or if age sets the orbital properties of stars irrespective of their $\alpha$-enhancement. Orbital actions $J_R$, $J_z$, and $J_\phi$ (or $L_z$) are our labels of stellar dynamics. We use ages for 58,278 LAMOST stars (measured to a precision of 40\%) within $\leq$2kpc of the Sun and we calculate orbital actions from proper motions and parallaxes given by Gaia's DR2. We find that \emph{at all ages}, the high- and low-$\alpha$ sequences are dynamically distinct. This implies separate formation and evolutionary histories for the two sequences; a star's membership in the high- or low-$\alpha$ sequence indicates its dynamical properties at a given time. We use action space to make an efficient selection of halo stars and subsequently report a group of old, low-$\alpha$ stars in the halo, which may be a discrete population from an infall event.
\end{abstract}

\bibstyle{aasjournal.bst}


\section{Introduction} \label{sec:intro}
% \todo[inline,color=green!20]{- Introduce high and low alpha sequence (e.g. Hayden et al., 2015)}
% -talk about spectroscopic surveys in general, and then LAMOST and Gaia.
% -why does $\alpha$-abundance help in studying galaxies, "galactic archeology"?\\

A growing number of extensive stellar surveys including GALAH \citep{GALAH2015}, LAMOST \citep{LAMOST2012,LAMOSTDR2-2016}, APOGEE \citep{APOGEE2017}, RAVE \citep{RAVE-DR5-2017,RAVE-on2017}, and Gaia \citep{Gaia2012,GaiaDR2} are cataloging larger regions of our Galaxy. Subsequently, understanding the structure and formation of the Milky Way is within our reach more than it has ever been before. Each part of the Milky Way (the bulge, disk, and halo) has stars which have different chemical and dynamical properties, as well as possibly different birth properties. Deducing the formation histories of various parts of the Milky Way by studying the chemistry, dynamics and ages of stars in different regions can lead us to understand epochs in the life of the Galaxy itself \citep[e.g.][]{Silva2018}. The disk of the Milky Way is the most expansive stellar component, where most of the stellar mass resides, and potentially holds immense information, which can shed light on how the universe evolves on galactic and cosmological scales.

Disk stars in the Milky Way appear to fall into two broad categories of chemical composition: (1) $\alpha$-enhanced or high-$\alpha$ stars, and (2) $\alpha$-poor or low-$\alpha$ stars \citep{Fuhrmann1998, Gratton2000, Prochaska2000, Bensby2003, Venn2004, Bensby2005, Adibekyan2011, Bensby&F&O-2014, Anders2014, Nidever2014, Hayden2015}. The $\alpha$-abundance measurement is typically derived from some ratio of elements fused in Helium capture (i.e. Mg, Si, Ti, Ca). The bimodality seen in the [Fe/H]-[$\alpha$-Fe] plane has been inferred to be the signature of two distinct populations \citep[e.g.][]{Mackereth2018, Clarke2019}, and chemical enrichment models suggest this requires a mix of populations with distinct enrichment histories \citep[e.g.][]{Nidever2015}. Spectroscopic surveys have established correlations between $\alpha$-enhancement and properties such as age \citep{Wyse&Gimlore1988,Haywood2013, Bensby&F&O-2014,Bergemann2014}: high-$\alpha$ stars are mostly old, and low-$\alpha$ stars are typically young. Quite possibly, the distinct formation timescales and velocity dispersion of the ``thin" and ``thick" components of the Milky Way disk could translate into $\alpha$-abundance signatures. However, the $\alpha$-enhancement of stars does not directly map to their assignment to the thin or thick disk \citep{Feltzing2008, Schonrich2009b, Adibekyan2011, Loebman2011, Bensby&F&O-2014, Silva2018,Bland-H2018}. Chemistry is our parameter of Galactic inquisition and we wish to understand how the high- and low-$\alpha$ sequences are distributed dynamically at a given age. We have the expectation that this will enable us to better understand the origin of the dichotomy that is seen in the [$\alpha$/Fe]-[Fe/H] plane. That is, if the birth and evolution tracks of the high- and low-$\alpha$ sequences might be discrete or related.

To make any sort of generalizable, empirical statement about how the two $\alpha$-sequences separate along other axes, we need access to accurately measured properties for a large sample of stars. Gaia DR2  \citep{GaiaDR2} measured positions, distances, and proper motions for $\sim$1.3 billion objects in the sky, which makes it possible to study the dynamics of stars with unprecedented accuracy. The Large sky Area Multi-Object Spectroscopic Telescope (LAMOST) Survey \citep{LAMOST2012} is one of the most extensive catalogs of low-resolution (R$\sim1800$) spectroscopic data at present. It is an optical ($3650-9000$\AA) survey, and its second data release \citep{LAMOSTDR2-2016} covered $\sim500,000$ red giants. \emph{The Cannon}, developed by \citet{Ness2015}, provides a data-driven method to infer information about a large number of stars by channeling and combining data from across various astronomical surveys. It has been a challenge to accurately and precisely measure ages for large samples of bright red giant stars that can be observed to vast distances across the Galaxy. This is because there is little discriminating power in stellar models in this regime. However, using \emph{The Cannon}, \citet{Ho2017ages} created the largest catalog of age estimates for $\sim230,000$ red giants by inferring the masses from [C/M] and [N/M] abundances. Data-driven modeling using the correlations between [C/M], [N/M], [Fe/H], and age (from asteroseismic determinations) has given us access to a far larger sample of stars for which we can determine ages \citep{Martig2016, Masseron2016, Ness2016}.
%talk about this sentence Thurs: It has been a challenge to accurately measure ages of stars because age can usually only be inferred from visible properties, such as chemical abundance from spectra \citep{Rix2013,Bovy2016,Ho2017ages}.

Actions provide a powerful description of stellar dynamics because they can encapsulate information to uniquely define an orbit \citep{Trick2018}. We use cylindrical coordinates $(r, \phi, z)$ for positions of stars and calculate actions using:
$$J_i \equiv \oint_{orbit} p_i dx_i$$
where $i= r, \phi, z$, and $p_i$ is the conjugate momentum \citep[for a detailed introduction to actions see ][]{Sellwood2014, binney-tremaine2008}. If we approximate the Milky Way's potential to be axisymmetric and its evolution over time to be slow enough, actions are constants of motion for timescales of the order of several orbits \citep[e.g.][]{Bland-H2018}. The three actions, $J_R$, $J_z$, and $J_{\phi}$ (or angular momentum, $L_z$) separately provide physical interpretations of a star's orbit. (We will use $L_z$ instead of $J_{\phi}$ in our discussion hereafter.) $J_R$ indicates the eccentricity of an orbit, or an orbit's deviation from a circle. $L_z$ is a direct indicator of the radius of the orbit, and $J_z$ indicates the maximum vertical distance of an orbit from the Galactic plane.

This paper is organized as follows: in \S2, we describe the main properties of our data sample, and explain our process of analysis. In \S3, we present our main results, and further analysis of the dynamics of the two $\alpha$-sequences as a function of [Fe/H] (\S3.2) and spatial selection $|z|$ (\S3.3), as well as a comparison of our results with ages derived from Bayesian inference (\S3.4). We discuss our results in \S4 and point out the detection a sample of old, low-$\alpha$ stars in the halo which might be part of the Gaia Enceladus (\S4.2) \citep[e.g.][]{Helmi2018, Belokurov2018}, before concluding in \S5.

% \todo[inline, color=green!20]{- Introduce LAMOST survey\\}
% \todo[inline, color=green!20]{- Introduce actions as numbers that = orbits DONE?\\
% - Introduce the idea of using actions+chemistry (orbits+star formation) to test for population formation and discreteness -DONE? \\
% - Relevant: Beane et al., 2018, Silva Aguire et al., 2018.\\
% - Introduce the idea of thin and thick disk in solar neighbourhood i.e. Duong et al. ,2018 -DONE? }

\section{Data}
\begin{figure*}[htbp]
  \centering
  \includegraphics[width=1\linewidth]{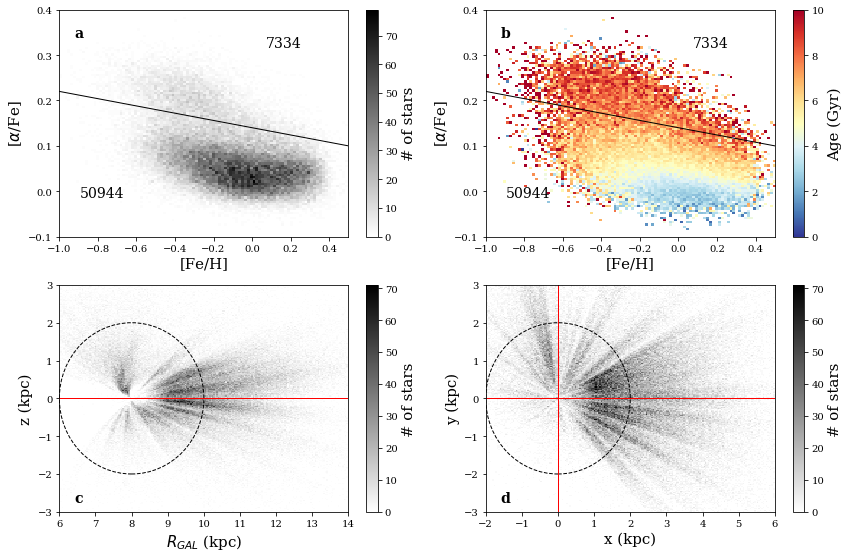}
%   \begin{tabular}[b]{c}
%     \includegraphics[width=1\linewidth]{figures/AM-vs-FeH-numdensity-age_edit1.png} \\
%     \small (a) \qquad \qquad \qquad \qquad \qquad \qquad \qquad \qquad \qquad \qquad \qquad \qquad \qquad \qquad(b)
%   \end{tabular}  \qquad
%       \begin{tabular}[b]{c}
%     \includegraphics[width=1\linewidth]{spatial-dist_edit1.png} \\
% \small (c) \qquad \qquad \qquad \qquad \qquad \qquad \qquad \qquad \qquad \qquad \qquad \qquad \qquad (d)  \end{tabular} \qquad
  \caption{Main properties of our data sample of the LAMOST disk red giants. (a) The number density distribution of $58,278$ stars is shown in the [$\alpha$/Fe]-[Fe/H] plane. We use the black line to differentiate between high- and low-$\alpha$ stars. There are 7,334 high-$\alpha$ stars and 50,944 low-$\alpha$ stars (as indicated in the plot). (b) The same plot as (a), but colored by age. We can see that high-$\alpha$ stars (above the black line) are mostly old, whereas the low-$\alpha$ stars (below the line) are typically much younger. (c) The spatial spread of the $\sim150,000$ strong sample is shown in a density plot of $z$ vs. Galactic radius ($R_{GAL}$). The Galactic plane at $z=0$ is shown by the horizontal red line. Our analyses are restricted to stars within $\leq$2kpc of the Sun, which are encompassed by the dashed circle. (d) The $x$-$y$ distribution of stars, again showing our restriction on distance, within the dashed circle of radius $\leq$ 2kpc. The intersection of the two red lines marks the position of the Sun, at $R_{GAL}$ = 8 kpc.}
  \label{fig:dataset}
  \vspace{10pt}
 \end{figure*}

We investigate $\sim60,000$ red giant stars in the disk ($\leq2$kpc from the Sun) taken from the LAMOST Survey and use their ages as calculated by \cite{Ho2017ages} (precise up to 0.2 dex), as well as the proper motion and parallax information acquired from Gaia DR2.

Figure~\ref{fig:dataset}a shows the number density distribution of our sample in the [$\alpha$/Fe]-[Fe/H] plane. We see that, as expected, there are two collections of stars that emerge clearly in this plane (separated roughly by the black line, which we set to differentiate high- and low-$\alpha$ stars). We denote the high-$\alpha$ stars as those that fall above the black line, and the low-$\alpha$ stars are those that fall below the black line. There are 7,334 high-$\alpha$ stars and $50,944$ low-$\alpha$ stars in our sample. On average, for our sample, the high-$\alpha$ sequence has an overall lower metallicity ([Fe/H]) compared to the low-$\alpha$ sequence.
At a metallicity of [Fe/H]$\sim-$0.2, there is a ``knee" in the plot (Figure~\ref{fig:dataset}a) after which the disparity in the amount of $\alpha$-elements and iron-peak elements goes down, and the high-$\alpha$ and low-$\alpha$ sequences appear to converge \citep[e.g.][]{Hayden2015}.

Figure~\ref{fig:dataset}b shows the distribution of our sample of (disk) stars once again in the [$\alpha$/Fe]-[Fe/H] plane, but colored by age. It becomes apparent that the mean ages of the two sequences are different, in the sense that high-$\alpha$ stars are typically older than the low-$\alpha$ population. However, there is overlap in the age distribution and both $\alpha$-sequences span stars across a broad range of ages.

Figures~\ref{fig:dataset}c,d show the spatial extent of the $\sim150,000$ LAMOST stars for which we have age information \citep[from ][]{Ho2017ages}.  This full sample extends from a Galactic radius of $\sim$6kpc to $\sim$14kpc, and $\sim$2kpc above and below the Galactic plane. However, we restrict our analysis to disk stars within a distance of $\leq$2kpc of the Sun. This distance cut was implemented because we have used inverse parallax as a proxy for distance, and that estimate becomes erroneous at large distances \citep{Bailer-Jones2018,Ting2018}. The spatial extent of stars that we take into account for our study are within the dashed circles in Figures~\ref{fig:dataset}c,d.

\begin{figure}[ht]
\centering
    \includegraphics[width=0.9\linewidth]{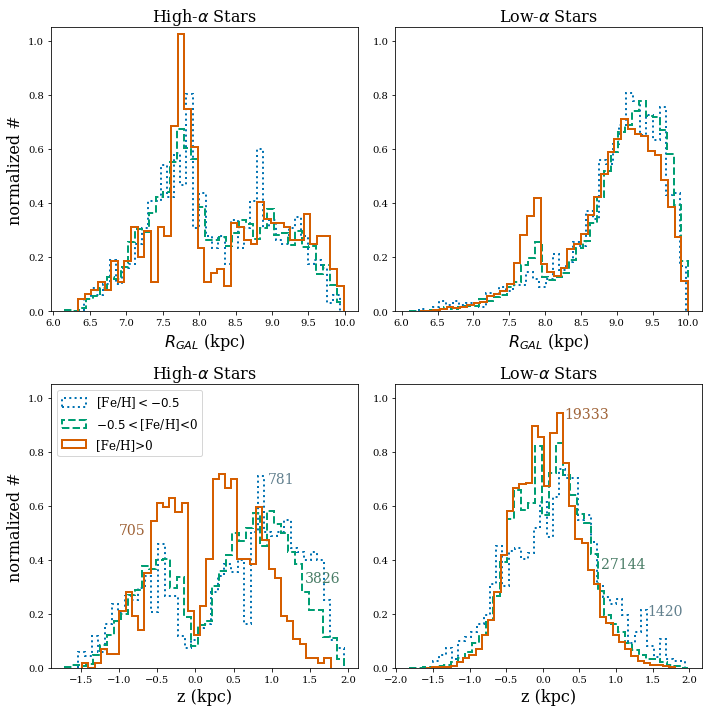} \\
\caption {Normalized histograms for $R_{GAL}$ (top panel) and $z$ (lower panel) for the two $\alpha$-sequences (high-$\alpha$ to the left, low-$\alpha$ to the right) in three different metallicity ranges: (i) [Fe/H] $<-0.5$ (light blue dotted line), (ii) $-0.5 < $[Fe/H]$<0$ (aquamarine dashed line), and (iii) [Fe/H] $> 0$ (orange solid line), as explained in the legend in the lower left subplot. The numbers of stars in each metallicity range and $\alpha$-sequence is indicated in the lower panels, in the respective color of the [Fe/H] range. We see in the top panel that the two $\alpha$-sequences have different mean radii for all metallicity ranges. The low-$\alpha$ stars are preferentially located at larger Galactic radii, compared to the low-$\alpha$ stars. In the lower panel, the dispersion of stars around the mid-plane is clearly much smaller for stars in the low-$\alpha$ sequence compared to the high-$\alpha$ sequence. However, within the  high-$\alpha$ sequence, the most metal-rich stars show the narrowest span in z.}
\label{fig:histograms-metallicity-bins}
\end{figure}

In Figure~\ref{fig:histograms-metallicity-bins}, the top panel shows normalized distributions of the Galactocentric radius, $R_{GAL}$, for the high- and low-$\alpha$ sequences, and the lower panel shows the spread of the distance from the Galactic plane, $z$, for three different metallicity ranges, (i) [Fe/H] $<-0.5$ (light blue dotted line), (ii) $-0.5 < $[Fe/H]$<0$ (aquamarine dashed line), and (iii) [Fe/H] $> 0$ (orange solid line), as explained in the legend in the lower left subplot. (This representation scheme is followed hereafter each time we categorize stars based on [Fe/H].) The numbers of stars in each metallicity range and $\alpha$-sequence is recorded in the bottom subplots in the respective color of the [Fe/H] range.
We see a weak gradient in the radii of low-$\alpha$ stars with respect to metallicity; mean [Fe/H] decreases at higher $R_{GAL}$ as expected from prior literature which has proposed that the radial metallicity gradient in the Milky Way comes about largely from low-$\alpha$ stars \citep[e.g.][]{Duong2018}.

% \todo[inline, color=green!20]{- quality cuts}
%Our sample has the following cuts as \cite{Ho2017ages} (\S5) which had enabled them to calculate ages for the stars accurate to 0.2 dex:
%$$ -0.8 < [Fe/H] < 0.25 $$
%$$4000 < T_{eff} < 5000 $$
%$$1.8 < log(g) < 3.3$$
%$$-0.05 < [\alpha/Fe] < 0.3$$
%\todo[inline,color=pink!40]{the lower limit on $[\alpha/Fe]$ above (0.05) doesn't seem consistent with fig. \ref{fig:dataset}a? Double checked with Anna's paper}

We have stellar ages precise to 0.2 dex from the carbon and nitrogen abundances catalogued in \cite{Ho2017ages}.  Additionally, in order to determine precise actions, we need precise and accurate distance measurements \citep[e.g.][]{Coronado2018,Trick2018,Beane2018}. We use the inverse of parallax $(\omega^{-1})$ as a proxy for distance and restrict our sample to stars which have a parallax error $(\delta\omega/\omega)$ of less than 20\%.

% \todo[inline, color=green!20]{- how did you calculate actions }
We calculate actions using the same method as \cite{Trick2018}: by first converting the observable positions and velocities into cartesian $(X, Y, Z)$ and heliocentric $(U_{HC}, V_{HC}, W_{HC})$ coordinates, and finally to galactocentric $(R,\phi,z,v_R,v_T,v_z)$ coordinates using the \galpy~package \citep{Bovy2015}. $U$ is the velocity component towards the Galactic center, $V$ is the component in the direction of rotation of the Galaxy, and $W$ is the component towards the Galactic north pole.
The coordinate conversion from cartesian and heliocentric to galactocentric is necessary to calculate the actions $J_R$, $J_z$, and $L_z$ in \galpy~and is done using $(U_\odot, V_\odot, W_\odot)=(11.1, 12.24, 7.25)$km $s^{-1}$ as the Sun's velocity in the Local Standard of Rest (LSR) \citep{Schonrich2010} and $(R_\odot,\phi_\odot,z_\odot)=(8,0,0.025)$ (with $R_\odot, z_\odot$ in kpc) as the Sun's position within the Galaxy \citep{Juric2008}.

%\todo[inline, color=pink!50]{typical errors on all inputs to action code\\
% }
% - [Fe/H] v [alpha/Fe] plane \\

\section{Results}

\subsection{Dynamical Actions as a Function of Age}
\begin{figure*}[ht]
  \centering
    \includegraphics[width=1\linewidth]{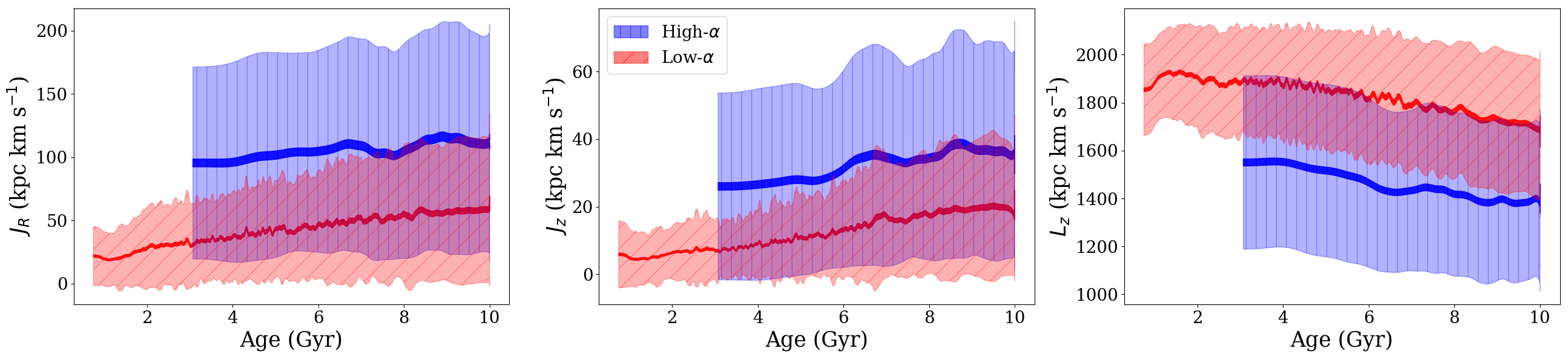} \\
    \caption{Key results of our work. They show that high- and low-$\alpha$ sequences display distinct dynamical trends at all ages, with significant scatter about the mean. The running means (with bin size of $N = 600$ stars) and standard deviations ($\sigma$) of each of the three actions for 7,334 high-$\alpha$ stars and 50,944 low-$\alpha$ stars are plotted as a function of age. The running means of high- and the low-$\alpha$ sequences are represented by the solid blue curve and the solid red curve respectively. The running means have been smoothed using a Gaussian kernel with $\sigma_s = 100$kpc km $s^{-1}$. The thickness of the solid curves represents the standard error ($\sigma / \sqrt{N-1}$), and the shaded regions are the respective 1-$\sigma$ standard deviation in actions (vertically hatched for high-$\alpha$ and diagonally hatched for low-$\alpha$).}
\label{fig:actions-vs-age}
  \end{figure*}

We examine the actions $J_R, J_z$, and $L_z$ for for 7,334 high-$\alpha$ stars and 50,944 low-$\alpha$ stars as a function of stellar age, and find that the high- and low-$\alpha$ stars have very distinct mean dynamical properties. Figure~\ref{fig:actions-vs-age} shows the running mean of each of the actions as a function of age for the two $\alpha$-sequences (high-$\alpha$ in blue and vertically hatched, low-$\alpha$ in red and diagonally hatched). In this Figure, each $\alpha$-sequence displays distinct dynamical behavior in all three orbital parameters: eccentricity ($J_R$), height above the Galactic plane ($J_z$), and radius ($L_z$). Moreover, the distinction in dynamical trends of the high- and low-$\alpha$ stars is present consistently at all ages.

% - explain how calculated running median/mean\\
The running means of actions and age were calculated using a bin size of $N=600$ stars for both the $\alpha$-sequences. The 1-$\sigma$ standard deviation ($\sigma$) is used to estimate the dispersion in the running means, and the standard errors are calculated as $\sigma/(\sqrt{N-1})$. There is a difference of almost a factor of 2 between the mean values of $J_R$ and $J_z$ for the high-$\alpha$ and low-$\alpha$ stars, with low-$\alpha$ stars having lower orbital eccentricities and lower vertical excursions. The low-$\alpha$ stars on average have $\sim1.25$ times larger angular momentum ($L_z$) than the high-$\alpha$ stars. There are slight gradients in the running means with respect to age, which might be affected by the selection function of the LAMOST survey. Since we have not taken the selection function into account, we make no quantitative measure or claims regarding the trend of the dynamics with respect to age for our sample. The dispersion around the running mean is significant and overlapping between the two $\alpha$-sequences.

In Table~\ref{tab:dispersions} we report the ratio of average dispersion ($\overline{\sigma}(J_i)$) to average action ($\bar{J}_i$) for each of the three actions of the two $\alpha$-sequences. These numbers reported only the average values, and they do not take into account the change in variance over time that is seen in $J_R$ and $J_z$ of low-$\alpha$ sequence. However, they do give an adequate idea of the relative dispersion in actions for the high- and the low-$\alpha$ stars in our sample.

% what is the variance in the running?
\begin{table}[h!]
\begin{center}
\begin{tabular}{ | m{1.02cm} | m{1.2cm} | m{1.2cm} | }
\hline
 \multirow{2}{4em}{Action} & \multicolumn{2}{c|}{$\overline{\sigma}(J_i)$ / $\bar{J}_i$} \\\cline{2-3}
 & high-$\alpha$ & low-$\alpha$ \\
 \hline
 $J_R$ & 0.79 & 0.98 \\
\hline
 $J_z$ & 0.9 & 1.13 \\
\hline
 $L_z$ & 0.24 & 0.13 \\
\hline
\end{tabular}
\end{center}
\caption{A table showing the mean 1-$\sigma$ dispersion ($\overline{\sigma}(J_i)$) divided by the mean of each action ($\bar{J}_i$), for the two $\alpha$-sequences. The dispersions and running means used in this table are those which are shown in Figure~\ref{fig:actions-vs-age}. \label{tab:dispersions}}
\end{table}

% - describe all figures explaining running mean, sampling error and 1-sigma standard deviation shown \\

% - how different are means?\\
%\todo[inline, color=green!20]{-does the variance in the running mean change as a function of age ?}

\subsection{Dynamical Actions for the High- and Low-$\alpha$ Sequence as a Function of [Fe/H] }

\begin{figure*}[htbp]
   \centering
%   \begin{tabular}[b]{c}
     \includegraphics[width=0.7\linewidth]{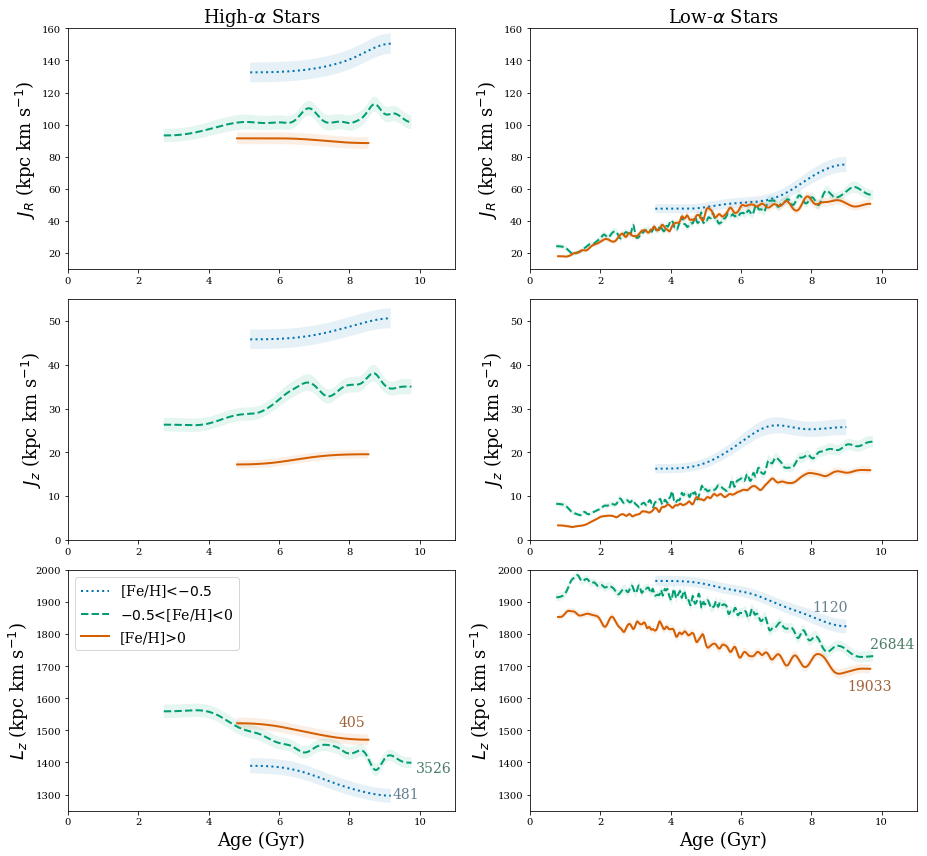}
%   \end{tabular} \qquad
  \caption{%\textbf{[need y axis labels at left (DONE)]}
  Smoothed running means of $J_R, J_z,$ and $L_z$ (with bin size $N =  300$) for the high-$\alpha$ (left panel) and low-$\alpha$ stars (right panel) as a function of age in three different metallicity ranges. Each [Fe/H] bin is represented by the same color scheme and line style as in Figure~\ref{fig:histograms-metallicity-bins}. The legend for all plots is once again shown in the bottom left subplot, and the number of stars in each [Fe/H] range and $\alpha$-sequence is recorded in the bottom panel in the respective color of the [Fe/H] range. We used a gaussian kernel with $\sigma_s = 100$kpc km $s^{-1}$ for smoothing the data. The standard errors ($\sigma/\sqrt{N-1}$) are shown by the shaded regions around each curve, which we can see are on the order of the thickness of the lines. These plots make it clear that the two $\alpha$-sequences have quite distinct dynamical properties (associated with each of the three actions)  at a given metallicity.}
  \label{fig: actions-in-metallicity-bins}
\end{figure*}

We now analyze the two $\alpha$-sequences as a function of [Fe/H] to examine if they also separate dynamically when conditioned on metallicity.
Figure~\ref{fig: actions-in-metallicity-bins} shows smoothed running means of $J_R, J_z,$ and $L_z$ versus age for high-$\alpha$ stars in the left panel, and the low-$\alpha$ stars in the right panel, for three different metallicity ranges of (i) [Fe/H] $ < -0.5$, (ii) $-0.5 <$[Fe/H]$ < 0$, and (iii) [Fe/H] $> 0$. The different [Fe/H] ranges are shown with different colors and line styles, as used in Figure~\ref{fig:histograms-metallicity-bins} and indicated in the legend. The standard errors on these curves are shown by the shaded regions.
From these plots, it is quite clear that the mean eccentricity, vertical excursion from the Galactic plane, and angular momentum of the two $\alpha$-sequences are distinct even if they have the same metallicity. This Figure also shows the richness of the information linking orbits to stellar chemistry. The high-$\alpha$ stars separate out significantly in $J_R$ compared to the low-$\alpha$ stars as a function of [Fe/H], with the lowest metallicity stars showing the highest radial actions. Similarly the high-$\alpha$ stars separate out more markedly, particularly at high metallicity, in $J_z$ compared to the low-$\alpha$ stars, as a function of [Fe/H], with the lowest metallicity stars showing the highest vertical actions. The two sequences are similarly dispersed in angular momentum ($L_z$) as a function of [Fe/H].

This examination demonstrates that there is clearly a relationship between stellar orbits and [Fe/H] that is determined by a star's membership in either the low and high-$\alpha$ sequences. While the stars separate dynamically for these two sequences, the metallicity of a star is additionally indicative of its dynamics at a given age, particularly for the high-$\alpha$ sequence.

% The distinction in $|z|$-distances of the two sequences is a little less clear from the histograms (partly because our sample is restricted to within 2$kpc$ from the plane), but appears to be prominent in the metal poor stars.

% \begin{figure}[htbp]
% \centering
%     \includegraphics[width=0.7\linewidth]{hi-lo-Fe-bins2kpc.png} \\
% \caption {Histograms for R- and $|z|$-distances for the two $\alpha$-sequences in three different metallicity bins. We see that the two sequences have quite distinguishable radii at all metallicities, whereas only the metal poor high- and low-$\alpha$ stars appear to have noticeably different $|z|$-values. We also see a slight indication of the radial metallicity gradient in the top right plot of R for low-$\alpha$ stars.}
% \label{fig:histograms-metallicity-bins}
% \end{figure}

\subsection{Dynamical Actions of the High- and Low-$\alpha$ Sequences as a Function of Spatial Selection, $|z|$}

\begin{figure*}[ht]
   \centering
    \includegraphics[width=0.7\linewidth]{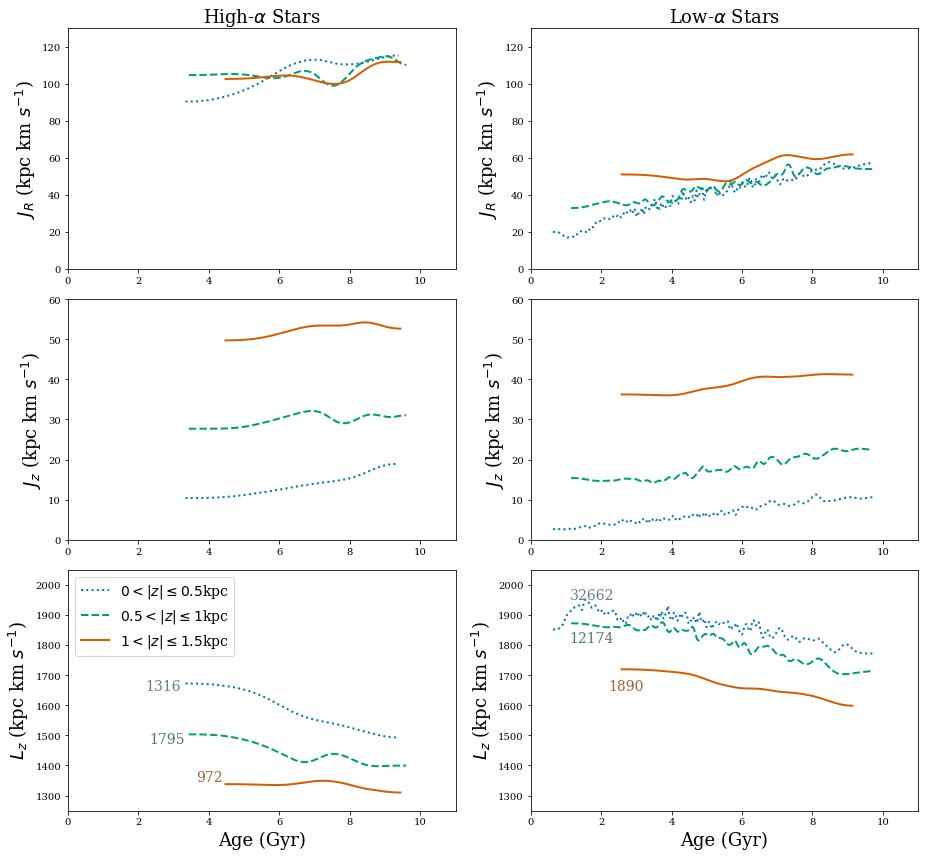}
  \caption{Smoothed running means of $J_R, J_z$, and $L_z$ versus age for the high- (left panel) and low-$\alpha$ (right panel) sequences separated into three ranges of height above the Galactic plane, $|z|$: (i) $0 < |z| \leq 0.5$kpc (light blue dotted line), (ii) $0.5 < |z| \leq 1$kpc (aquamarine dashed line), and (iii) $1 < |z| \leq 1.5$kpc (orange solid line). The legend is shown in the bottom left plot and number of stars in each $|z|$ range and $\alpha$-sequence are recorded in the bottom panel in the respective color of each $|z|$ range. A gaussian kernel with $\sigma_s = 100$kpc km $s^{-1}$ was used for smoothing the running means. Division into $|z|$ ranges helps decouple the $|z|$-positions of the stars from their chemistry and dynamics. We can see that even at the same height from the plane, the two $\alpha$-sequences have different $J_R, J_z,$ and $L_z$, i.e., different dynamical properties, at all ages. Standard errors have been left out in these plots because they are on the order of the thickness of the curves.}
 \label{fig:z-bins}
\end{figure*}

The measured actions are a function of spatial position of the stars; stars at higher distances from the plane in the disk have higher $J_z$ values and the high-$\alpha$ sequence is preferentially distributed at larger $|z|$, at the Sun. We therefore wish to examine in more detail the differences in the action distributions of the high- and low-$\alpha$ sequences conditioned on height from the plane ($|z|$). We examine this by testing the respective actions of the high- and low-$\alpha$ sequences in three narrow slices in $|z|$: (i) $0 < |z| \leq 0.5$kpc, (ii) $0.5 < |z| \leq 1$kpc, and (iii) $1 < |z| \leq 1.5$kpc.  We find that at the same height from the Galactic plane, the two $\alpha$-sequences have mean actions that are different from each other at all ages, as shown in Figure~\ref{fig:z-bins}. The standard errors on all curves (not shown in the Figure) are on the order of the thickness of the lines.

It can be seen that the high-$\alpha$ stars have significantly higher eccentricity ($J_R$) at all ages compared to the low-$\alpha$ stars even when the stars have the same $|z|$-distances. Angular momentum ($L_z$) as a function of age is much lower for high-$\alpha$ stars compared to the low-$\alpha$ stars (at a given $|z|$-distance), with stellar angular momenta decreasing with increasing distance from the Galactic plane. Note that there are small differences in $J_z$ as a function of age (middle panel of Figure~\ref{fig:z-bins}) for the two $\alpha$-sequences; this indicates that within our bins of height $|z|$, the low-$\alpha$ sequence stars are preferentially nearer to the Galactic plane.

These plots demonstrate that the orbital properties of the stars in each $\alpha$-sequence are sensitive to the spatial selection of the stars in height above the plane, which reflects the correlation between structure and chemistry in the Galaxy and also the selection function of the survey. However, it is clear that at a given spatial selection in height above the plane there is the clear separation in the mean of the dynamical action between the two $\alpha$-sequences.

\subsection{Comparison with Ages Derived from Bayesian Inference}
\begin{figure*}[htbp]
 \centering
    \begin{tabular}[b]{c}
    \includegraphics[width=1\linewidth]{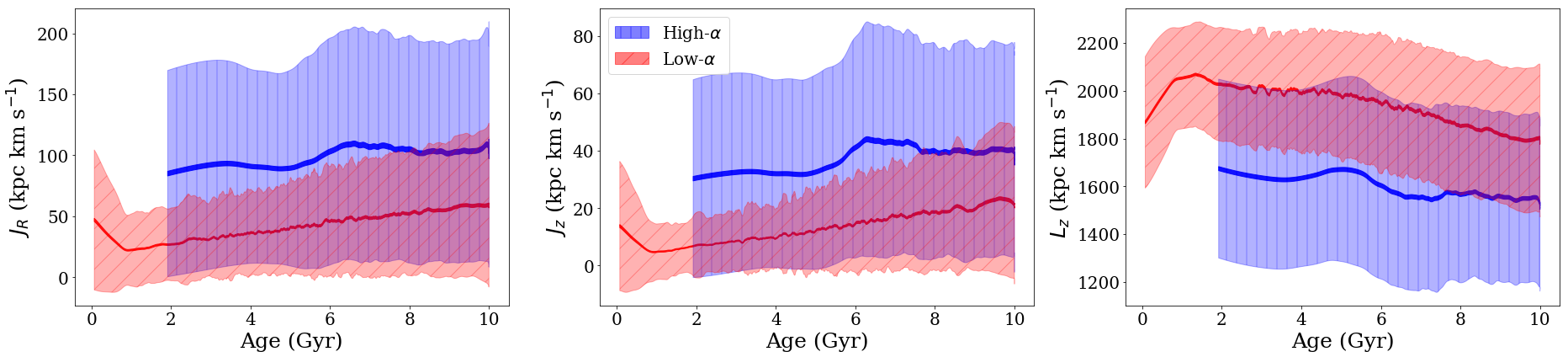}
  \end{tabular} \qquad
  \caption{The same plots as in Figure~\ref{fig:actions-vs-age} but using \citet{SD2018} ages and actions. The running means (with bin size $N = 2000$ stars) and standard deviations ($\sigma$) of each of the three actions for 30,579 high-$\alpha$ stars and 122,668 low-$\alpha$ stars are plotted as a function of age. The high- and the low-$\alpha$ sequences are represented by the solid blue curve and the solid red curve respectively. The running means have been smoothed with a gaussian kernel using $\sigma_s = 200$kpc km $s^{-1}$. The thickness of the solid curves represents the standard error ($\sigma/\sqrt{N-1}$), and the hatched shaded regions are the respective 1-$\sigma$ standard deviations in actions. These plots quite closely match Figure~\ref{fig:actions-vs-age}.}
  \label{fig: S&D}
\end{figure*}

  We also used ages and actions from a catalog developed by \citet{SD2018} to test whether or not we get similar results with their data. In their analysis, Sanders \& Das employed Bayesian neural networks to date stars, and the Gaia DR2 data to find the actions using the Stackel fudge method. Figure~\ref{fig: S&D} shows the plots obtained using the data from Sanders \& Das (2018), made in the same way as Figure~\ref{fig:actions-vs-age}. These plots have 30,579 high-$\alpha$ stars, and 122,668 low-$\alpha$ stars restricted to within $\leq2$kpc of the Sun, and we did a quality cut such that parallax error is $<20\%$. That these data also indicate a distinction in dynamical trends between the two $\alpha$-sequences is quite encouraging.
% \begin{figure}[h!]
%  \centering
%  \begin{tabular}[b]{c}
%     \includegraphics[width=.4\linewidth]{AM-vs-FeH-SD.png} \\
%   \end{tabular} \qquad \\
%   \caption{[Alpha/M] vs Metallicity for the Sanders\&Das (2018) disk ($<=2kpc$) stars colored by age}
% \end{figure}

%In Figure~\ref{fig:metallicity-bins} we can see that ...

\section{Discussion}
\subsection{Orbital properties of the High- and Low-$\alpha$ Sequences}
Studying the actions $J_R, J_z,$ and $L_z$ as a function of age for the two $\alpha$-sequences makes it clear that disk stars in the solar neighborhood that fall into different $\alpha$-enrichment groups have very different mean orbital properties at \emph{all} ages. The fact that this dichotomy in the action plane is present throughout all ages, for all actions, is a strong indication that high- and low-$\alpha$ sequences form distinct populations which have different evolution and birth properties. We also find that within a given $\alpha$-sequence, a star's [Fe/H] is additionally informative as to its orbital properties, particularly for $J_z$ and $J_R$ for the high-$\alpha$ sequence.
%If they were not distinct populations, we would expect the orbital properties to depend heavily on the fundamental parameter of age, and not chemical enrichment.
% - Interpretation of each action
Figure~\ref{fig:actions-vs-age} shows our main results, from which we can understand what the nature of the orbits of stars in the high- or low-$\alpha$ sequence is, at least with respect to each other. Orbits of high-$\alpha$ stars are on average more eccentric (have greater $J_R$) than the low-$\alpha$ stars, almost consistently by a factor of two. We see a similar relationship between the two sequences in how much their orbits diverge from the Galactic plane (or $J_z$): the high-$\alpha$ sequence has orbits which go almost up to twice the height from the plane compared to low-$\alpha$ stars. The angular momentum ($L_z$) of low-$\alpha$ stars is $\sim1.25$ times that of the high-$\alpha$ stars, which indicates that low-$\alpha$ stars on average have larger radii than high-$\alpha$ stars \citep[e.g.][]{Beane2018, Hayden2015}. Our results are consistent with \citet{Mackereth2019}, who find similar relationships between $\alpha$-abundance and velocity dispersions ($\sigma_z$, $\sigma_R$) of high- and low-$\alpha$ stars. Our findings are also aligned with the analysis of \citet{Nidever2015}, who determine single population chemical evolution models to be insufficient to explain the $\alpha$-bimodality seen in the APOGEE data. These empirical findings can be explained in a cosmological context if the high-$\alpha$ sequence has a distinct, early and rapid star formation origin,  in ``clumps" \citep[e.g.][]{Clarke2019}. Such clumps are also observed at high redshift in young spiral Galaxies that are presumably typical Milky Way progenitors \citep{Guo2015}.

It is interesting to note that we find a valley between two slight peaks (in Figure~\ref{fig:actions-vs-age}) in $J_R$ and $J_z$ for the high-$\alpha$ sequence between age $\sim$6-9Gyr, and a slight peak between two valleys in $L_z$ for high-$\alpha$ stars in the same age window. We see similar signatures in analyzing LAMOST stars from \citet{SD2018}, as can be seen in Figure~\ref{fig: S&D}. The presence of a perturbation in $J_z$ was also suggested by \citet{Beane2018} (\S3.2) for stars in the APOKASC2 sample (independent of $\alpha$-enrichment). Such a perturbation could be a signature of an infall event in the Milky Way, which altered the dynamical properties of stars at $\sim6$ Gyr. For a bump in actions rather than a net step increase, such an infall would likely trigger the  formation of new stars (e.g., a gas cloud merger), as opposed to a merger with a star cluster which would bring in pre-existing stars.

%Had the merging object been a star cluster, it would have kicked the orbits of Milky Way stars of all ages, resulting in a sort of step in the actions in Figure~\ref{fig:actions-vs-age} after $\sim6$ Gyr, and not peaks localized in age. A gas cloud merger, on the other hand, would trigger formation of stars with $J_R, J_z$, and $L_z$ distinct from other pre-existing stars in the neighborhood, and the newly formed stars would all have approximately one age.

We note that these ``bumps'' are localized in $\alpha$-enrichment as well as metallicity, which means that any infalling gas cloud would have a consistent average chemical composition. As can be seen in Figure~\ref{fig: actions-in-metallicity-bins}, they appear only in the high-$\alpha$ sequence in the range of $-0.5<$[Fe/H]$<0$, within the same age range ($\sim$6-9 Gyr). This leads us to predict that we might expect differences in the detailed elemental composition of this possibly foreign population compared to other high-$\alpha$ stars in the Galaxy of this age.

\subsection{Discovery of Old and $\alpha$-Poor Halo Stars in the LAMOST Sample}

Numerous works \citep[e.g.][]{Silva2018, Ness2016,  Bensby&F&O-2014, Bergemann2014, Haywood2013, Adibekyan2011} have established that high-$\alpha$ stars of the disk are old and low-$\alpha$ stars of the disk are typically young. Furthermore, Milky Way halo stars ($\sim$3-5kpc from the Sun) are typically high-$\alpha$, metal poor ([Fe/H]$ < -1.0$), and old (e.g. \citet{Freudenburg2017,Bovy2012, Lee2011, Prochaska2000, Fuhrmann1998}. We now examine the $\alpha$-sequences for the halo stars in LAMOST including stars within $\leq5$kpc of the Sun.

\begin{figure*}[htbp]
\centering
\begin{tabular}[b]{c}
\includegraphics[width=1\linewidth]{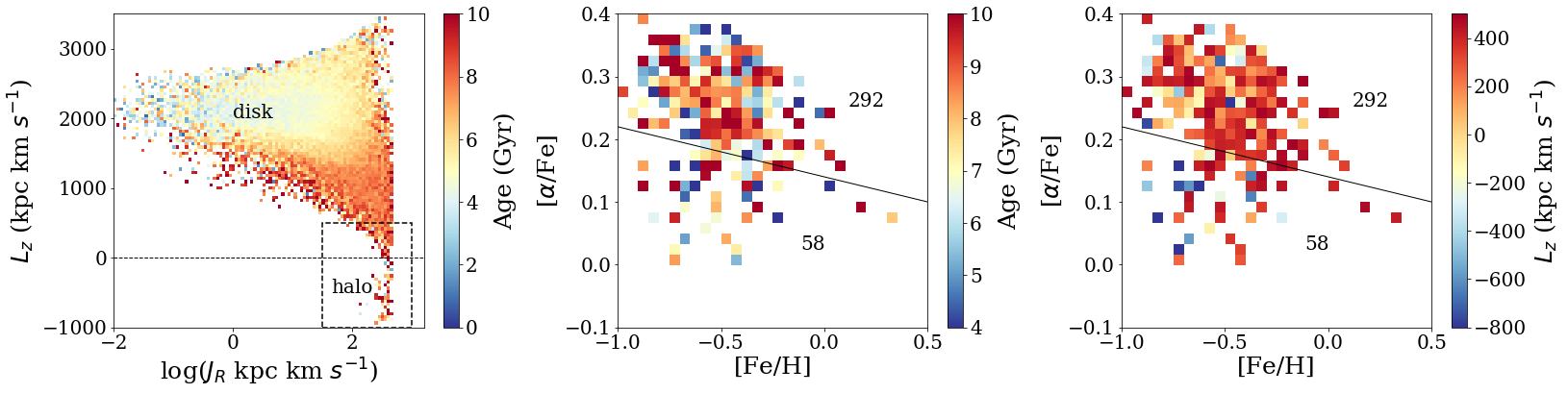} \\ \small (a) \qquad \qquad \qquad \qquad \qquad  \qquad  \qquad  \qquad  (b) \qquad  \qquad  \qquad  \qquad  \qquad  \qquad  \qquad  \qquad (c)
\end{tabular} \qquad
  \caption{(a) The distribution of $144,231$ high- and low-$\alpha$ stars within $\leq$5kpc from the Sun in the $L_z$-log$_{10}(J_R)$ plane, colored by age. The $L_z$-log$_{10}(J_R)$ plane visually separates the disk and the halo; stellar orbits with a variety of angular momenta ($L_z$) and eccentricity ($J_R$) values represent the disk, resulting in the triangular shape of the distribution, whereas the tapering bottom end of the distribution with low $L_z$ and high $J_R$ characterizes the halo orbits. In the adjacent plots, we examine the $\alpha$-enhancement, age, and orbital motion of halo stars (within the dashed box) with log$_{10}(J_R) > 1.5$ and $L_z < 500$kpc km $s^{-1}$.  (b) The distribution of 350 stars within 5kpc of the Sun with $L_z < 500$kpc km $s^{-1}$ and log$_{10}(J_R) > 1.5$ are shown in the [$\alpha$/Fe]-[Fe/H] plane, colored by age. The numbers of high- and low-$\alpha$ stars are indicated above and below the black separation line respectively. We find that the 58 low-$\alpha$ stars that make this cut have a mean age of $\sim$7.2Gyr, with a 1-$\sigma$ standard deviation of $\sim$2.5Gyr, whereas the 292 high-$\alpha$ stars which make this cut have a mean age of $\sim$7.8Gyr with a 1-$\sigma$ standard deviation of $\sim$2.3Gyr. (c) The same as (b), but colored by angular momentum ($L_z$). Of the 58 low-$\alpha$ stars, 21 exhibit retrograde motion, i.e., have $L_z < 0$.}
\label{fig:infall-analysis}
\end{figure*}

%There is also a distinction between the two $\alpha$-sequences according to where within the disk they lie: the high-$\alpha$ sequence is believed to reside in the thick disk, whereas the low-$\alpha$ sequence is mostly in the thin disk of the galaxy \citep{Haywood2013}.
To select halo stars in our sample, we take stars that have high radial eccentricities ($J_R$) and low angular momenta ($L_z$). Figure~\ref{fig:infall-analysis}a shows how the LAMOST stars are distributed across the $L_z$-log$_{10}(J_R)$ plane; the disk flares out and appears joined to the halo which is distributed around an $L_z \sim 0$ and high $J_z$, as indicated in the box drawn around this region. Such a selection in actions discriminates halo stars from the disk population. This discrete cut is not intended as an absolute marker of halo membership but these stars have orbital properties which align with halo membership and are quite distinct from the disk \citep[e.g.][]{Helmi2018}.

We now show the distribution of our selected halo stars in Figures~\ref{fig:infall-analysis}b,c coloured by age and angular momentum, respectively. We use the same line as previously to differentiate high- and low-$\alpha$ stars. Surprisingly, we find low-$\alpha$ stars in the halo in our sample. There are 292 high-$\alpha$ stars and 58 low-$\alpha$ stars which make the halo cut. Respectively, the age distribution of the high- and low-$\alpha$ sequences is similar, with a mean age and 1-$\sigma$ standard deviation of (7.8,2.3)Gyr for  the high-$\alpha$ halo stars and (7.2,2.5)Gyr for the low-$\alpha$ stars. These ages may be badly measured, but we found no indication of this in examining the $\chi^2$ ($\overline{\chi^2} \sim 0.5$) and age uncertainties ($\overline{\Delta_{age}} \sim 2.2$Gyr) on these stars.
%We show the distribution in the $[Fe/H]-[\alpha/Fe]$ plane for these 350 high- and low-$\alpha$ stars with $L_z < 500~kpc~km~s^{-1}$ and $J_R$ $> 150~kpc~km~s^{-1}$ in fig.~\ref{fig:infall-analysis}a (colored by age), and fig.~\ref{fig:infall-analysis}b (colored by $L_z$).

Out of the 58 low-$\alpha$ halo stars, 21 exhibit retrograde motion with $L_z < 0$. The metallicity distribution of mean [Fe/H] and the standard error of the mean ($\sigma_{err}=\sigma/\sqrt{N-1}$) is different for the retrograde compared to the prograde low-$\alpha$ stars. The retrograde stars have ($\overline{[Fe/H]}\pm\sigma_{err}$) $=(-0.70\pm0.08)$, whereas it is $(-0.45\pm0.04)$ for the prograde stars. The retrograde stars are also younger (than the prograde stars) with ($\overline{age}\pm\sigma_{err}) =(6.23\pm0.59)$Gyr, whereas for the prograde stars it is $(7.75\pm0.39)$Gyr.
%\textbf{We need to do a mock error analysis, given errors of 40 percent - take a Gaussian distribution of stars centered on the mean age of the low alpha sequence and generate 1000 samples and see how many fall above 7 GYrs, we want to verify this number that we see that are old is not just due to the error distribution) - will explain on Friday}.
%happen to be atypically older than the rest of the low-$\alpha$ sequence.
% \todo[inline, color=pink!40]{we currently don't have a figure showing this population of old low-$\alpha$ stars}
%Following Gaia DR2, a spurt of research has found strong indications of prior mergers with the Milky Way

This anomalous low-$\alpha$ halo population may be associated with the coherent population of infall discovered in the Gaia data by a number of groups \citep{Myeong2018, Koppelman2018,Helmi2018}. In particular, \citet{Helmi2018} find a retrograde population in the halo within 5kpc of the Sun which they call the ``Gaia Enceladus" and show that it might have merged with the Milky Way $\sim10$ Gyr ago.
At [Fe/H]$\sim-0.6$, the Gaia Enceladus was demonstrated to be photometrically consistent with an age range of $\sim$10-13 Gyr \citep{Helmi2018}.  This infall
includes low-$\alpha$ stars as demonstrated using the APOGEE data in \citet{Helmi2018}, however has not been previously spectroscopically age dated.

If the stars in the LAMOST data that we have identified as potential Gaia Enceladus members are ex-situ to the Milky Way, they will have experienced a different star formation rate and therefore  have different birth and evolution history from the typical stars in the Milky Way. So we expect multiple abundances to show anomalous trends in other elements, which can be tested using individual abundance derivations from LAMOST data \citep{Belokurov2018, Ting2017}. Additional dimensions of information along these lines will enable us to determine if these stars are from a single or multiple set of infall events.

%The fact that the old, low-$\alpha$ stars seen by \citet{Helmi2018} belong in the halo implies that their orbits must be highly eccentric for them to be observed in the solar neighborhood, and thus $J_R$ for these stars must be higher on average, and their average radii (or $L_z$) can be expected to be less than the majority of stars in our sample. To confirm whether the subpopulation of old, low-$\alpha$ stars that we see coincides with the stars in Gaia Enceladus, we examine the dynamics of the low-$\alpha$ sequence by plotting $L_z$ vs $J_R$ colored by age. The $L_z$ vs $J_R$ plane in Fig.\ref{fig:old-low-alpha-in-halo} separates the disk and halo stars in the following way: stars with a wide range of eccentricity ($J_R$) and angular momenta ($L_z$) represent the disk, whereas low angular momentum and high eccentricity are indicative of halo orbits. The figure clarifies that the old low-$\alpha$ stars in our sample do indeed lie near the halo. Therefore, it is highly probable that this atypical subpopulation of old low-$\alpha$ stars in our sample can at least partially be explained by the Gaia Enceladus.

\section{Conclusion}
We calculated actions ($J_R, J_z,$ and $L_z$) using dynamical information from Gaia DR2, and used ages for LAMOST red giants (in the disk near the Sun) from \citet{Ho2017ages}. We have found clear indications that $\alpha$-enrichment is correlated with dynamical properties of stellar orbits. Our main conclusions are summarized below:
\begin{enumerate}
\item Consistently and at all ages, the high- and low-$\alpha$ sequences exhibit distinct mean trends in eccentricity ($J_R$), divergence from Galactic plane ($J_z$), and angular momentum ($L_z$)---the three dynamical properties which (simplistically, but uniquely) characterize a stellar orbit. This result can be seen clearly in Figure~\ref{fig:actions-vs-age}. This implies that the two $\alpha$-sequences are distinct populations in the Milky Way, with different birth and evolution properties. This conclusion is also supported by studies which use models to simulate Milky Way-like galaxies and find that the two $\alpha$-sequences emerge via distinct birth and evolutionary tracks \citep[e.g.][]{Mackereth2018EAGLE, Grand2018}.
\item The conclusion that chemistry is strongly linked to orbital properties of stars at all ages has been shown to be consistent at a given  metallicity (Figure~\ref{fig: actions-in-metallicity-bins}) as well as at a narrow slice in height from the plane (Figure~\ref{fig:z-bins}).
\item The $L_z$-log$_{10}(J_R)$ plane is an effective space to select for halo stars. We find a population of low-$\alpha$ stars which are atypically older and live predominantly in the halo, which may be remnants of merger events in the Milky Way, e.g. Gaia Enceladus \citep{Helmi2018, Belokurov2018}.
\end{enumerate}
Our findings of high- and low-$\alpha$ sequences being dynamically distinct at all ages, including at a given ([Fe/H], $|z|$) is strong evidence for the discrete origin of these populations in building the disk. These findings are consistent at different spatial locations and for different stellar surveys with different selection functions (e.g. APOGEE \citep{Mackereth2019, Nidever2014}).
\section{Acknowledgements}
We would like to thank the Flatiron Institute and Columbia University for providing the infrastructure and support for this research. We also thank Glennys Farrar, Angus Beane, and David Hogg for helpful conversations.

\bibliography{main}  %%edit made here

\begin{thebibliography}{}
\expandafter\ifx\csname natexlab\endcsname\relax\def\natexlab#1{#1}\fi
\providecommand{\url}[1]{\href{#1}{#1}}

\bibitem[{{Adibekyan} {et~al.}(2011){Adibekyan}, {Santos}, {Sousa}, \&
  {Israelian}}]{Adibekyan2011}
{Adibekyan}, V.~Z., {Santos}, N.~C., {Sousa}, S.~G., \& {Israelian}, G. 2011,
  \aap, 535, L11

\bibitem[{{Anders} {et~al.}(2014){Anders}, {Chiappini}, {Santiago},
  {Rocha-Pinto}, {Girardi}, {da Costa}, {Maia}, {Steinmetz}, {Minchev},
  {Schultheis}, {Boeche}, {Miglio}, {Montalb{\'a}n}, {Schneider}, {Beers},
  {Cunha}, {Allende Prieto}, {Balbinot}, {Bizyaev}, {Brauer}, {Brinkmann},
  {Frinchaboy}, {Garc{\'{\i}}a P{\'e}rez}, {Hayden}, {Hearty}, {Holtzman},
  {Johnson}, {Kinemuchi}, {Majewski}, {Malanushenko}, {Malanushenko},
  {Nidever}, {O'Connell}, {Pan}, {Robin}, {Schiavon}, {Shetrone}, {Skrutskie},
  {Smith}, {Stassun}, \& {Zasowski}}]{Anders2014}
{Anders}, F., {Chiappini}, C., {Santiago}, B.~X., {et~al.} 2014, \aap, 564,
  A115

\bibitem[{{Bailer-Jones} {et~al.}(2018){Bailer-Jones}, {Rybizki}, {Fouesneau},
  {Mantelet}, \& {Andrae}}]{Bailer-Jones2018}
{Bailer-Jones}, C.~A.~L., {Rybizki}, J., {Fouesneau}, M., {Mantelet}, G., \&
  {Andrae}, R. 2018, \aj, 156, 58

\bibitem[{{Beane} {et~al.}(2018){Beane}, {Ness}, \& {Bedell}}]{Beane2018}
{Beane}, A., {Ness}, M., \& {Bedell}, M. 2018, ArXiv e-prints, arXiv:1807.05986

\bibitem[{{Belokurov} {et~al.}(2018){Belokurov}, {Erkal}, {Evans}, {Koposov},
  \& {Deason}}]{Belokurov2018}
{Belokurov}, V., {Erkal}, D., {Evans}, N.~W., {Koposov}, S.~E., \& {Deason},
  A.~J. 2018, \mnras, 478, 611

\bibitem[{{Bensby} {et~al.}(2003){Bensby}, {Feltzing}, \&
  {Lundstr{\"o}m}}]{Bensby2003}
{Bensby}, T., {Feltzing}, S., \& {Lundstr{\"o}m}, I. 2003, \aap, 410, 527

\bibitem[{{Bensby} {et~al.}(2005){Bensby}, {Feltzing}, {Lundstr{\"o}m}, \&
  {Ilyin}}]{Bensby2005}
{Bensby}, T., {Feltzing}, S., {Lundstr{\"o}m}, I., \& {Ilyin}, I. 2005, \aap,
  433, 185

\bibitem[{{Bensby} {et~al.}(2014){Bensby}, {Feltzing}, \&
  {Oey}}]{Bensby&F&O-2014}
{Bensby}, T., {Feltzing}, S., \& {Oey}, M.~S. 2014, \aap, 562, A71

\bibitem[{{Bergemann} {et~al.}(2014){Bergemann}, {Ruchti}, {Serenelli},
  {Feltzing}, {Alves-Brito}, {Asplund}, {Bensby}, {Gruyters}, {Heiter},
  {Hourihane}, {Korn}, {Lind}, {Marino}, {Jofre}, {Nordlander}, {Ryde},
  {Worley}, {Gilmore}, {Randich}, {Ferguson}, {Jeffries}, {Micela},
  {Negueruela}, {Prusti}, {Rix}, {Vallenari}, {Alfaro}, {Allende Prieto},
  {Bragaglia}, {Koposov}, {Lanzafame}, {Pancino}, {Recio-Blanco}, {Smiljanic},
  {Walton}, {Costado}, {Franciosini}, {Hill}, {Lardo}, {de Laverny}, {Magrini},
  {Maiorca}, {Masseron}, {Morbidelli}, {Sacco}, {Kordopatis}, \& {Tautvai{\v
  s}ien{\.e}}}]{Bergemann2014}
{Bergemann}, M., {Ruchti}, G.~R., {Serenelli}, A., {et~al.} 2014, \aap, 565,
  A89

\bibitem[{Binney \& Tremaine(2008)}]{binney-tremaine2008}
Binney, J., \& Tremaine, S. 2008, Princeton University Press, Princeton, NJ,
  23, 389

\bibitem[{{Bland-Hawthorn} {et~al.}(2018){Bland-Hawthorn}, {Sharma},
  {Tepper-Garcia}, {Binney}, {Freeman}, {Hayden}, {Kos}, {De Silva}, {Lewis},
  {Asplund}, {Buder}, {Casey}, {D'Orazi}, {Duong}, {Lin}, {Lind}, {Martell},
  {Ness}, {Simpson}, {Zucker}, {Zwitter}, {Kafle}, {Quillen}, {Ting}, \&
  {Wyse}}]{Bland-H2018}
{Bland-Hawthorn}, J., {Sharma}, S., {Tepper-Garcia}, T., {et~al.} 2018, ArXiv
  e-prints, arXiv:1809.02658

\bibitem[{{Bovy}(2015)}]{Bovy2015}
{Bovy}, J. 2015, \apjs, 216, 29

\bibitem[{Bovy {et~al.}(2012)Bovy, Rix, \& Hogg}]{Bovy2012}
Bovy, J., Rix, H.-W., \& Hogg, D.~W. 2012, The Astrophysical Journal, 751, 131

\bibitem[{{Casey} {et~al.}(2017){Casey}, {Hawkins}, {Hogg}, {Ness}, {Rix},
  {Kordopatis}, {Kunder}, {Steinmetz}, {Koposov}, {Enke}, {Sanders}, {Gilmore},
  {Zwitter}, {Freeman}, {Casagrande}, {Matijevi{\v c}}, {Seabroke},
  {Bienaym{\'e}}, {Bland-Hawthorn}, {Gibson}, {Grebel}, {Helmi}, {Munari},
  {Navarro}, {Reid}, {Siebert}, \& {Wyse}}]{RAVE-on2017}
{Casey}, A.~R., {Hawkins}, K., {Hogg}, D.~W., {et~al.} 2017, \apj, 840, 59

\bibitem[{{Clarke} {et~al.}(2019){Clarke}, {Debattista}, {Nidever}, {Loebman},
  {Simons}, {Kassin}, {Du}, {Ness}, {Fisher}, {Quinn}, {Wadsley}, {Freeman}, \&
  {Popescu}}]{Clarke2019}
{Clarke}, A.~J., {Debattista}, V.~P., {Nidever}, D.~L., {et~al.} 2019, \mnras,
  484, 3476

\bibitem[{{Coronado} {et~al.}(2018){Coronado}, {Rix}, \&
  {Trick}}]{Coronado2018}
{Coronado}, J., {Rix}, H.-W., \& {Trick}, W.~H. 2018, ArXiv e-prints,
  arXiv:1804.07760

\bibitem[{{Cui} {et~al.}(2012){Cui}, {Zhao}, {Chu}, {Li}, {Li}, {Zhang}, {Su},
  {Yao}, {Wang}, {Xing}, {Li}, {Zhu}, {Wang}, {Gu}, {Luo}, {Xu}, {Zhang},
  {Liu}, {Zhang}, {Yang}, {Cao}, {Chen}, {Chen}, {Chen}, {Chen}, {Chu}, {Feng},
  {Gong}, {Hou}, {Hu}, {Hu}, {Hu}, {Jia}, {Jiang}, {Jiang}, {Jiang}, {Jin},
  {Li}, {Li}, {Li}, {Liu}, {Liu}, {Lu}, {Mao}, {Men}, {Qi}, {Qi}, {Shi},
  {Tang}, {Tao}, {Wang}, {Wang}, {Wang}, {Wang}, {Wang}, {Wang}, {Wang},
  {Wang}, {Wang}, {Wang}, {Wang}, {Wang}, {Xu}, {Xu}, {Yang}, {Yu}, {Yuan},
  {Yuan}, {Zhai}, {Zhang}, {Zhang}, {Zhang}, {Zhao}, {Zhou}, {Zhou}, {Zhu}, \&
  {Zou}}]{LAMOST2012}
{Cui}, X.-Q., {Zhao}, Y.-H., {Chu}, Y.-Q., {et~al.} 2012, Research in Astronomy
  and Astrophysics, 12, 1197

\bibitem[{{De Silva} {et~al.}(2015){De Silva}, {Freeman}, {Bland-Hawthorn},
  {Martell}, {de Boer}, {Asplund}, {Keller}, {Sharma}, {Zucker}, {Zwitter},
  {Anguiano}, {Bacigalupo}, {Bayliss}, {Beavis}, {Bergemann}, {Campbell},
  {Cannon}, {Carollo}, {Casagrande}, {Casey}, {Da Costa}, {D'Orazi}, {Dotter},
  {Duong}, {Heger}, {Ireland}, {Kafle}, {Kos}, {Lattanzio}, {Lewis}, {Lin},
  {Lind}, {Munari}, {Nataf}, {O'Toole}, {Parker}, {Reid}, {Schlesinger},
  {Sheinis}, {Simpson}, {Stello}, {Ting}, {Traven}, {Watson}, {Wittenmyer},
  {Yong}, \& {{\v Z}erjal}}]{GALAH2015}
{De Silva}, G.~M., {Freeman}, K.~C., {Bland-Hawthorn}, J., {et~al.} 2015,
  \mnras, 449, 2604

\bibitem[{{Duong} {et~al.}(2018){Duong}, {Freeman}, {Asplund}, {Casagrande},
  {Buder}, {Lind}, {Ness}, {Bland-Hawthorn}, {De Silva}, {D'Orazi}, {Kos},
  {Lewis}, {Lin}, {Martell}, {Schlesinger}, {Sharma}, {Simpson}, {Zucker},
  {Zwitter}, {Anguiano}, {Da Costa}, {Hyde}, {Horner}, {Kafle}, {Nataf},
  {Reid}, {Stello}, {Ting}, \& {Wyse}}]{Duong2018}
{Duong}, L., {Freeman}, K.~C., {Asplund}, M., {et~al.} 2018, \mnras, 476, 5216

\bibitem[{{Feltzing} \& {Bensby}(2008)}]{Feltzing2008}
{Feltzing}, S., \& {Bensby}, T. 2008, Physica Scripta Volume T, 133, 014031

\bibitem[{{Freudenburg} {et~al.}(2017){Freudenburg}, {Weinberg}, {Hayden}, \&
  {Holtzman}}]{Freudenburg2017}
{Freudenburg}, J.~K.~C., {Weinberg}, D.~H., {Hayden}, M.~R., \& {Holtzman},
  J.~A. 2017, \apj, 849, 17

\bibitem[{{Fuhrmann}(1998)}]{Fuhrmann1998}
{Fuhrmann}, K. 1998, \aap, 338, 161

\bibitem[{{Gaia Collaboration} {et~al.}(2018){Gaia Collaboration}, {Brown},
  {Vallenari}, {Prusti}, {de Bruijne}, {Babusiaux}, {Bailer-Jones}, {Biermann},
  {Evans}, {Eyer}, \& et~al.}]{GaiaDR2}
{Gaia Collaboration}, {Brown}, A.~G.~A., {Vallenari}, A., {et~al.} 2018, \aap,
  616, A1

\bibitem[{{Gilmore} {et~al.}(2012){Gilmore}, {Randich}, {Asplund}, {Binney},
  {Bonifacio}, {Drew}, {Feltzing}, {Ferguson}, {Jeffries}, {Micela}, \&
  et~al.}]{Gaia2012}
{Gilmore}, G., {Randich}, S., {Asplund}, M., {et~al.} 2012, The Messenger, 147,
  25

\bibitem[{{Grand} {et~al.}(2018){Grand}, {Bustamante}, {G{\'o}mez}, {Kawata},
  {Marinacci}, {Pakmor}, {Rix}, {Simpson}, {Sparre}, \& {Springel}}]{Grand2018}
{Grand}, R.~J.~J., {Bustamante}, S., {G{\'o}mez}, F.~A., {et~al.} 2018, \mnras,
  474, 3629

\bibitem[{{Gratton} {et~al.}(2000){Gratton}, {Carretta}, {Matteucci}, \&
  {Sneden}}]{Gratton2000}
{Gratton}, R.~G., {Carretta}, E., {Matteucci}, F., \& {Sneden}, C. 2000, \aap,
  358, 671

\bibitem[{{Guo} {et~al.}(2015){Guo}, {Ferguson}, {Bell}, {Koo}, {Conselice},
  {Giavalisco}, {Kassin}, {Lu}, {Lucas}, {Mandelker}, {McIntosh}, {Primack},
  {Ravindranath}, {Barro}, {Ceverino}, {Dekel}, {Faber}, {Fang}, {Koekemoer},
  {Noeske}, {Rafelski}, \& {Straughn}}]{Guo2015}
{Guo}, Y., {Ferguson}, H.~C., {Bell}, E.~F., {et~al.} 2015, \apj, 800, 39

\bibitem[{{Hayden} {et~al.}(2015){Hayden}, {Bovy}, {Holtzman}, {Nidever},
  {Bird}, {Weinberg}, {Andrews}, {Majewski}, {Allende Prieto}, {Anders},
  {Beers}, {Bizyaev}, {Chiappini}, {Cunha}, {Frinchaboy},
  {Garc{\'{\i}}a-Her{\'n}andez}, {Garc{\'{\i}}a P{\'e}rez}, {Girardi},
  {Harding}, {Hearty}, {Johnson}, {M{\'e}sz{\'a}ros}, {Minchev}, {O'Connell},
  {Pan}, {Robin}, {Schiavon}, {Schneider}, {Schultheis}, {Shetrone},
  {Skrutskie}, {Steinmetz}, {Smith}, {Wilson}, {Zamora}, \&
  {Zasowski}}]{Hayden2015}
{Hayden}, M.~R., {Bovy}, J., {Holtzman}, J.~A., {et~al.} 2015, \apj, 808, 132

\bibitem[{{Haywood} {et~al.}(2013){Haywood}, {Di Matteo}, {Lehnert}, {Katz}, \&
  {G{\'o}mez}}]{Haywood2013}
{Haywood}, M., {Di Matteo}, P., {Lehnert}, M.~D., {Katz}, D., \& {G{\'o}mez},
  A. 2013, \aap, 560, A109

\bibitem[{{Helmi} {et~al.}(2018){Helmi}, {Babusiaux}, {Koppelman}, {Massari},
  {Veljanoski}, \& {Brown}}]{Helmi2018}
{Helmi}, A., {Babusiaux}, C., {Koppelman}, H.~H., {et~al.} 2018, ArXiv
  e-prints, arXiv:1806.06038

\bibitem[{{Ho} {et~al.}(2017){Ho}, {Rix}, {Ness}, {Hogg}, {Liu}, \&
  {Ting}}]{Ho2017ages}
{Ho}, A.~Y.~Q., {Rix}, H.-W., {Ness}, M.~K., {et~al.} 2017, \apj, 841, 40

\bibitem[{{Juri{\'c}} {et~al.}(2008){Juri{\'c}}, {Ivezi{\'c}}, {Brooks},
  {Lupton}, {Schlegel}, {Finkbeiner}, {Padmanabhan}, {Bond}, {Sesar},
  {Rockosi}, {Knapp}, {Gunn}, {Sumi}, {Schneider}, {Barentine}, {Brewington},
  {Brinkmann}, {Fukugita}, {Harvanek}, {Kleinman}, {Krzesinski}, {Long},
  {Neilsen}, {Nitta}, {Snedden}, \& {York}}]{Juric2008}
{Juri{\'c}}, M., {Ivezi{\'c}}, {\v Z}., {Brooks}, A., {et~al.} 2008, \apj, 673,
  864

\bibitem[{{Koppelman} {et~al.}(2018){Koppelman}, {Helmi}, \&
  {Veljanoski}}]{Koppelman2018}
{Koppelman}, H., {Helmi}, A., \& {Veljanoski}, J. 2018, \apjl, 860, L11

\bibitem[{{Kunder} {et~al.}(2017){Kunder}, {Kordopatis}, {Steinmetz},
  {Zwitter}, {McMillan}, {Casagrande}, {Enke}, {Wojno}, {Valentini},
  {Chiappini}, {Matijevi{\v c}}, {Siviero}, {de Laverny}, {Recio-Blanco},
  {Bijaoui}, {Wyse}, {Binney}, {Grebel}, {Helmi}, {Jofre}, {Antoja}, {Gilmore},
  {Siebert}, {Famaey}, {Bienaym{\'e}}, {Gibson}, {Freeman}, {Navarro},
  {Munari}, {Seabroke}, {Anguiano}, {{\v Z}erjal}, {Minchev}, {Reid},
  {Bland-Hawthorn}, {Kos}, {Sharma}, {Watson}, {Parker}, {Scholz}, {Burton},
  {Cass}, {Hartley}, {Fiegert}, {Stupar}, {Ritter}, {Hawkins}, {Gerhard},
  {Chaplin}, {Davies}, {Elsworth}, {Lund}, {Miglio}, \&
  {Mosser}}]{RAVE-DR5-2017}
{Kunder}, A., {Kordopatis}, G., {Steinmetz}, M., {et~al.} 2017, \aj, 153, 75

\bibitem[{{Lee} {et~al.}(2011){Lee}, {Beers}, {An}, {Ivezi{\'c}}, {Just},
  {Rockosi}, {Morrison}, {Johnson}, {Sch{\"o}nrich}, {Bird}, {Yanny},
  {Harding}, \& {Rocha-Pinto}}]{Lee2011}
{Lee}, Y.~S., {Beers}, T.~C., {An}, D., {et~al.} 2011, \apj, 738, 187

\bibitem[{{Loebman} {et~al.}(2011){Loebman}, {Ro{\v s}kar}, {Debattista},
  {Ivezi{\'c}}, {Quinn}, \& {Wadsley}}]{Loebman2011}
{Loebman}, S.~R., {Ro{\v s}kar}, R., {Debattista}, V.~P., {et~al.} 2011, \apj,
  737, 8

\bibitem[{{Luo} {et~al.}(2016){Luo}, {Zhao}, {Zhao}, {Deng}, {Liu}, {Jing},
  {Wang}, {Zhang}, {Shi}, {Cui}, {Chu}, {Li}, {Bai}, {Wu}, {Cai}, {Cao}, {Cao},
  {Carlin}, {Chen}, {Chen}, {Chen}, {Chen}, {Chen}, {Chen}, {Chen},
  {Christlieb}, {Chu}, {Cui}, {Dong}, {Du}, {Fan}, {Feng}, {Fu}, {Gao}, {Gong},
  {Gu}, {Guo}, {Han}, {He}, {Hou}, {Hou}, {Hou}, {Hu}, {Hu}, {Hu}, {Huo},
  {Jia}, {Jiang}, {Jiang}, {Jiang}, {Jin}, {Kong}, {Kong}, {Lei}, {Li}, {Li},
  {Li}, {Li}, {Li}, {Li}, {Li}, {Li}, {Li}, {Li}, {Li}, {Li}, {Liang}, {Lin},
  {Liu}, {Liu}, {Liu}, {Liu}, {Lu}, {Luo}, {Mao}, {Newberg}, {Ni}, {Qi}, {Qi},
  {Shen}, {Shi}, {Song}, {Song}, {Su}, {Su}, {Tang}, {Tao}, {Tian}, {Wang},
  {Wang}, {Wang}, {Wang}, {Wang}, {Wang}, {Wang}, {Wang}, {Wang}, {Wang},
  {Wang}, {Wang}, {Wang}, {Wang}, {Wang}, {Wang}, {Wang}, {Wang}, {Wang},
  {Wang}, {Wei}, {Wei}, {Wu}, {Wu}, {Wu}, {Wu}, {Xing}, {Xu}, {Xu}, {Xu},
  {Yan}, {Yang}, {Yang}, {Yang}, {Yang}, {Yao}, {Yu}, {Yuan}, {Yuan}, {Yuan},
  {Yuan}, {Zhai}, {Zhang}, {Zhang}, {Zhang}, {Zhang}, {Zhang}, {Zhang},
  {Zhang}, {Zhang}, {Zhao}, {Zhou}, {Zhou}, {Zhu}, {Zhu}, {Zou}, \&
  {Zuo}}]{LAMOSTDR2-2016}
{Luo}, A.-L., {Zhao}, Y.-H., {Zhao}, G., {et~al.} 2016, VizieR Online Data
  Catalog, 5149

\bibitem[{{Mackereth} {et~al.}(2018{\natexlab{a}}){Mackereth}, {Crain},
  {Schiavon}, {Schaye}, {Theuns}, \& {Schaller}}]{Mackereth2018}
{Mackereth}, J.~T., {Crain}, R.~A., {Schiavon}, R.~P., {et~al.}
  2018{\natexlab{a}}, \mnras, 477, 5072

\bibitem[{{Mackereth} {et~al.}(2018{\natexlab{b}}){Mackereth}, {Crain},
  {Schiavon}, {Schaye}, {Theuns}, \& {Schaller}}]{Mackereth2018EAGLE}
---. 2018{\natexlab{b}}, \mnras, 477, 5072

\bibitem[{{Mackereth} {et~al.}(2019){Mackereth}, {Bovy}, {Leung}, {Schiavon},
  {Trick}, {Chaplin}, {Cunha}, {Feuillet}, {Majewski}, {Martig}, {Miglio},
  {Nidever}, {Pinsonneault}, {Silva Aguirre}, {Sobeck}, {Tayar}, \&
  {Zasowski}}]{Mackereth2019}
{Mackereth}, J.~T., {Bovy}, J., {Leung}, H.~W., {et~al.} 2019, arXiv e-prints,
  arXiv:1901.04502

\bibitem[{{Majewski} {et~al.}(2017){Majewski}, {Schiavon}, {Frinchaboy},
  {Allende Prieto}, {Barkhouser}, {Bizyaev}, {Blank}, {Brunner}, {Burton},
  {Carrera}, {Chojnowski}, {Cunha}, {Epstein}, {Fitzgerald}, {Garc{\'{\i}}a
  P{\'e}rez}, {Hearty}, {Henderson}, {Holtzman}, {Johnson}, {Lam}, {Lawler},
  {Maseman}, {M{\'e}sz{\'a}ros}, {Nelson}, {Nguyen}, {Nidever}, {Pinsonneault},
  {Shetrone}, {Smee}, {Smith}, {Stolberg}, {Skrutskie}, {Walker}, {Wilson},
  {Zasowski}, {Anders}, {Basu}, {Beland}, {Blanton}, {Bovy}, {Brownstein},
  {Carlberg}, {Chaplin}, {Chiappini}, {Eisenstein}, {Elsworth}, {Feuillet},
  {Fleming}, {Galbraith-Frew}, {Garc{\'{\i}}a}, {Garc{\'{\i}}a-Hern{\'a}ndez},
  {Gillespie}, {Girardi}, {Gunn}, {Hasselquist}, {Hayden}, {Hekker}, {Ivans},
  {Kinemuchi}, {Klaene}, {Mahadevan}, {Mathur}, {Mosser}, {Muna}, {Munn},
  {Nichol}, {O'Connell}, {Parejko}, {Robin}, {Rocha-Pinto}, {Schultheis},
  {Serenelli}, {Shane}, {Silva Aguirre}, {Sobeck}, {Thompson}, {Troup},
  {Weinberg}, \& {Zamora}}]{APOGEE2017}
{Majewski}, S.~R., {Schiavon}, R.~P., {Frinchaboy}, P.~M., {et~al.} 2017, \aj,
  154, 94

\bibitem[{{Martig} {et~al.}(2016){Martig}, {Fouesneau}, {Rix}, {Ness},
  {M{\'e}sz{\'a}ros}, {Garc{\'{\i}}a-Hern{\'a}ndez}, {Pinsonneault},
  {Serenelli}, {Silva Aguirre}, \& {Zamora}}]{Martig2016}
{Martig}, M., {Fouesneau}, M., {Rix}, H.-W., {et~al.} 2016, \mnras, 456, 3655

\bibitem[{{Masseron} {et~al.}(2016){Masseron}, {Merle}, \&
  {Hawkins}}]{Masseron2016}
{Masseron}, T., {Merle}, T., \& {Hawkins}, K. 2016, {BACCHUS: Brussels
  Automatic Code for Characterizing High accUracy Spectra}, Astrophysics Source
  Code Library, , , ascl:1605.004

\bibitem[{{Myeong} {et~al.}(2018){Myeong}, {Evans}, {Belokurov}, {Sanders}, \&
  {Koposov}}]{Myeong2018}
{Myeong}, G.~C., {Evans}, N.~W., {Belokurov}, V., {Sanders}, J.~L., \&
  {Koposov}, S.~E. 2018, \apjl, 856, L26

\bibitem[{{Ness} {et~al.}(2015){Ness}, {Hogg}, {Rix}, {Ho}, \&
  {Zasowski}}]{Ness2015}
{Ness}, M., {Hogg}, D.~W., {Rix}, H.-W., {Ho}, A.~Y.~Q., \& {Zasowski}, G.
  2015, \apj, 808, 16

\bibitem[{{Ness} {et~al.}(2016){Ness}, {Hogg}, {Rix}, {Martig}, {Pinsonneault},
  \& {Ho}}]{Ness2016}
{Ness}, M., {Hogg}, D.~W., {Rix}, H.-W., {et~al.} 2016, \apj, 823, 114

\bibitem[{{Nidever} {et~al.}(2014){Nidever}, {Bovy}, {Bird}, {Andrews},
  {Hayden}, {Holtzman}, {Majewski}, {Smith}, {Robin}, {Garc{\'{\i}}a
  P{\'e}rez}, {Cunha}, {Allende Prieto}, {Zasowski}, {Schiavon}, {Johnson},
  {Weinberg}, {Feuillet}, {Schneider}, {Shetrone}, {Sobeck},
  {Garc{\'{\i}}a-Hern{\'a}ndez}, {Zamora}, {Rix}, {Beers}, {Wilson},
  {O'Connell}, {Minchev}, {Chiappini}, {Anders}, {Bizyaev}, {Brewington},
  {Ebelke}, {Frinchaboy}, {Ge}, {Kinemuchi}, {Malanushenko}, {Malanushenko},
  {Marchante}, {M{\'e}sz{\'a}ros}, {Oravetz}, {Pan}, {Simmons}, \&
  {Skrutskie}}]{Nidever2014}
{Nidever}, D.~L., {Bovy}, J., {Bird}, J.~C., {et~al.} 2014, \apj, 796, 38

\bibitem[{{Nidever} {et~al.}(2015){Nidever}, {Holtzman}, {Allende Prieto},
  {Beland}, {Bender}, {Bizyaev}, {Burton}, {Desphande}, {Fleming},
  {Garc{\'{\i}}a P{\'e}rez}, {Hearty}, {Majewski}, {M{\'e}sz{\'a}ros}, {Muna},
  {Nguyen}, {Schiavon}, {Shetrone}, {Skrutskie}, {Sobeck}, \&
  {Wilson}}]{Nidever2015}
{Nidever}, D.~L., {Holtzman}, J.~A., {Allende Prieto}, C., {et~al.} 2015, \aj,
  150, 173

\bibitem[{{Prochaska} {et~al.}(2000){Prochaska}, {Naumov}, {Carney},
  {McWilliam}, \& {Wolfe}}]{Prochaska2000}
{Prochaska}, J.~X., {Naumov}, S.~O., {Carney}, B.~W., {McWilliam}, A., \&
  {Wolfe}, A.~M. 2000, \aj, 120, 2513

\bibitem[{{Sanders} \& {Das}(2018)}]{SD2018}
{Sanders}, J.~L., \& {Das}, P. 2018, \mnras, 481, 4093

\bibitem[{{Sch{\"o}nrich} \& {Binney}(2009)}]{Schonrich2009b}
{Sch{\"o}nrich}, R., \& {Binney}, J. 2009, \mnras, 399, 1145

\bibitem[{{Sch{\"o}nrich} {et~al.}(2010){Sch{\"o}nrich}, {Binney}, \&
  {Dehnen}}]{Schonrich2010}
{Sch{\"o}nrich}, R., {Binney}, J., \& {Dehnen}, W. 2010, \mnras, 403, 1829

\bibitem[{{Sellwood}(2014)}]{Sellwood2014}
{Sellwood}, J.~A. 2014, Reviews of Modern Physics, 86, 1

\bibitem[{{Silva Aguirre} {et~al.}(2018){Silva Aguirre}, {Bojsen-Hansen},
  {Slumstrup}, {Casagrande}, {Kawata}, {Ciuc{\v a}}, {Handberg}, {Lund},
  {Mosumgaard}, {Huber}, {Johnson}, {Pinsonneault}, {Serenelli}, {Stello},
  {Tayar}, {Bird}, {Cassisi}, {Hon}, {Martig}, {Nissen}, {Rix},
  {Sch{\"o}nrich}, {Sahlholdt}, {Trick}, \& {Yu}}]{Silva2018}
{Silva Aguirre}, V., {Bojsen-Hansen}, M., {Slumstrup}, D., {et~al.} 2018,
  \mnras, 475, 5487

\bibitem[{{Ting} \& {Rix}(2018)}]{Ting2018}
{Ting}, Y.-S., \& {Rix}, H.-W. 2018, arXiv e-prints, arXiv:1808.03278

\bibitem[{{Ting} {et~al.}(2017){Ting}, {Rix}, {Conroy}, {Ho}, \&
  {Lin}}]{Ting2017}
{Ting}, Y.-S., {Rix}, H.-W., {Conroy}, C., {Ho}, A.~Y.~Q., \& {Lin}, J. 2017,
  \apjl, 849, L9

\bibitem[{{Trick} {et~al.}(2018){Trick}, {Coronado}, \& {Rix}}]{Trick2018}
{Trick}, W.~H., {Coronado}, J., \& {Rix}, H.-W. 2018, ArXiv e-prints,
  arXiv:1805.03653

\bibitem[{{Venn} {et~al.}(2004){Venn}, {Irwin}, {Shetrone}, {Tout}, {Hill}, \&
  {Tolstoy}}]{Venn2004}
{Venn}, K.~A., {Irwin}, M., {Shetrone}, M.~D., {et~al.} 2004, \aj, 128, 1177

\bibitem[{{Wyse} \& {Gilmore}(1988)}]{Wyse&Gimlore1988}
{Wyse}, R.~F.~G., \& {Gilmore}, G. 1988, \aj, 95, 1404

\end{thebibliography}

\end{document}